\documentclass[12pt]{iopart}

\newcommand{\bra}[2] {\mbox{}_{#2}\langle #1 |}
\newcommand{\ket}[2] {| #1 \rangle_{#2}}

\newcommand{\ee}[1] {\mathrm{e}^{#1}}

\newcommand{\dg}{^{\dagger}}

\usepackage{iopams}  
\usepackage{graphicx}
\usepackage{color}

\begin{document}

\title{Self-calibrating Quantum State Tomography}

\author{Agata M. Bra\'nczyk, Dylan H. Mahler, Lee A. Rozema, Ardavan Darabi, Aephraim M. Steinberg, Daniel F. V. James}
\address{CQIQC and IOS, Department of Physics, University of Toronto, 60 Saint George St., Toronto, Ontario M5S 1A7, 
Canada}

\begin{abstract}
We introduce and experimentally demonstrate a technique for performing quantum state tomography on multiple-qubit states despite  incomplete knowledge about the unitary operations used to change the measurement basis. Given unitary operations with \emph{unknown} rotation angles, our method can be used to reconstruct the density matrix of the state up to local $\hat\sigma_z$ rotations as well as recover the magnitude of the unknown rotation angle. We demonstrate high-fidelity \emph{self-calibrating tomography} on polarization-encoded one- and two-photon states. The unknown unitary operations are realized in two ways: using a birefringent polymer sheet---an inexpensive smartphone screen protector---or alternatively a liquid crystal wave plate with a tuneable retardance. We explore how our technique may be adapted for quantum state tomography of systems such as biological molecules where the magnitude and orientation of the transition dipole moment is not known with high accuracy. 

\end{abstract}

\maketitle
\section{Introduction}
Quantum state characterization is essential to the development of  quantum technologies, such as quantum computing \cite{Kok2007,Childs2010}, quantum information \cite{Nielsen2000} and quantum cryptography \cite{Bennett1984}. The measurement of multiple copies of the quantum state and subsequent reconstruction of the state's density matrix is known as \emph{quantum state tomography} (QST) \cite{James2001,Altepeter2004}. 

QST relies on performing measurements in a number of different bases to gain access to all the information required to reconstruct the density matrix. Different measurement bases are typically obtained by applying unitary operations to the state before measurement. For systems that couple to the electromagnetic field via their transition dipole, precise knowledge of the magnitude and orientation of the dipole moment is vital to performing such measurement-basis changes.  For example, QST of trapped ion qubits requires precise control of the orientation of the ions via magnetic fields \cite{Benhelm2008}. 

These techniques can not be immediately applied to systems that are either not well characterized, such as biological molecules, e.g. those found in photosynthetic systems \cite{Engel2007,Collini2010}, where the magnitude and orientation of the transition dipole moment is not known with high accuracy; or systems that experience variability during the fabrication process, such as colloidal quantum dots \cite{Guyot-Sionnest2008}. 

With such systems in mind we introduce a method for \emph{self-calibrating tomography} (SCT), which successfully reconstructs the state of the system despite incomplete knowledge about the unitary operations used to change the measurement basis. Other efforts to characterise photosynthetic systems are also underway, e.g. quantum process tomography using ultrafast spectroscopy was proposed by Yuen-Zhou \emph{et al.} \cite{Yuen-Zhou2011c,Yuen-Zhou2011b} and Hamiltonian tomography was proposed by Maruyama \emph{et al.} \cite{Maruyama2011}.

Our method for self-calibrating tomography may also improve the robustness of standard tomography, where calibrated unitary operations undergo small errors and fluctuations over the course of the experiment. For other modifications to standard QST due to inaccessible information or preferable measurements choices, we refer the reader to references \cite{Altepeter2005,Ling2006,Adamson2007,Rehacek2007,deBurgh2008,Adamson2008,Paris2009,Adamson2010,Nunn2010,Bogdanov2010,Medendorp2011, Yamagata2011, Brida2011,Asorey2011,Bogdanov2011,Teo2011,Teo2012}. 

In Section \ref{sec:theory}, we show that given local unitary operations  with unknown rotation angles, but known and adjustable rotation axes, it is possible to reconstruct the density matrix of a state up to local $\hat\sigma_z$ rotations, as well as recover the magnitude of the unknown rotation angles.

We demonstrate SCT in a linear-optical system using polarized photons as qubits in Section \ref{sec:exp}. An inexpensive smartphone screen protector, i.e. an uncharacterized birefringent polymer sheet, is used to change the measurement basis. We go on to investigate the technique's robustness to measurement noise and retardance magnitude, and demonstrate SCT of a two-qubit state using liquid crystal wave plates with tuneable retardances. 

In section \ref{sec:photo}, we investigate the application of this technique to quantum state tomography of photosynthetic systems.

\section{Theory}\label{sec:theory}

In this section, we give a brief review of quantum state tomography and introduce our technique for self-calibrating tomography.

\subsection{Quantum State Tomography}

The state of a qubit, given by the density matrix $\hat\rho$, can be decomposed into a sum of orthogonal operators $\sigma_{i}$:
\begin{eqnarray}\label{eq:1}
\hat\rho=\frac{1}{2}\sum_{i=0}^{3}\lambda_{i}\hat\sigma_{i}\,,
\end{eqnarray} 
where $\hat\sigma_0$ is the identity operator and $\hat\sigma_{1\mathrm{-}3}$ are the Pauli matrices. The coefficients $\lambda_{j}$ are given by the expectation values of the orthogonal operators, $\lambda_{i}=\mathrm{Tr}[\hat\rho\hat\sigma_{i}]$. 

Given an unknown state, $\hat\rho$ can be reconstructed by performing a number of projective measurements $\hat\mu_{j}$ on subsequent copies of the state. The measurement statistics are given by  
\begin{equation}\label{eq:1b}
n_{j}=\mathcal{N}_{j}\mathrm{Tr}[\hat\rho\hat\mu_{j}]\,,
\end{equation}
where $\mathcal{N}_{j}$ is a constant that depends on the duration of data collection, detector efficiency, loss etc.  Inserting (\ref{eq:1}) into (\ref{eq:1b}) gives
\begin{eqnarray}\label{eq:2}
n_{j}=\frac{\mathcal{N}_{j}}{2}\sum_{i=0}^{3}  \lambda_{i}\mathrm{Tr}[\hat\mu_{j}\hat\sigma_{i}]\,.
\end{eqnarray}
The equations given by (\ref{eq:2}) can be solved for $\lambda_{i}$, which can then be inserted into (\ref{eq:1}) to reconstruct the unknown density matrix $\hat\rho$. The operators $\hat\mu_{j}$ must be chosen such that the equations in (\ref{eq:2}) form a linearly independent set. If $\mathcal{N}_{j}$ are known, three different measurements suffice, given the normalization condition $\lambda_0=1$. Typically, $\mathcal{N}_{j}$ is unknown, but assumed to be equivalent for all measurements, in which case, four measurements are required to reconstruct the state. In practice, due to fluctuations in measurement statistics, it is advantageous to employ a maximum likelihood estimation method \cite{James2001} to reconstruct $\hat\rho$ rather than directly solving (\ref{eq:1b}), which can lead to unphysical density matrices.

We can define the measurement operators $\hat\mu_{j}$ in terms of a fixed known measurement operator $\hat\mu_{0}$ and different unitary operations $\hat U_{j}$ as follows.
\begin{equation}
\hat\mu_{j}=\hat U_{j}\dg\hat\mu_0\hat U_{j}\,.
\end{equation}
In terms of the measurement statistics, this is equivalent to applying the unitary operator $\hat U_{j}$ to the state prior to making measurements given by $\hat\mu_{0}$:
\begin{equation}
n_{j}=\mathcal{N}_{j}\mathrm{Tr}[\hat\rho\hat U_{j}\dg\hat\mu_0\hat U_{j}]=\mathcal{N}_{j}\mathrm{Tr}[\hat U_{j}\hat\rho\hat U_{j}\dg\hat\mu_0]\,,
\end{equation}
where we used the cyclic properties of the trace. 

\subsection{Self-calibrating Tomography}

Now consider the case where the unitary operator, used to change the measurement basis, is characterized by an unknown parameter $\alpha$. Recall that a unitary operation is simply a rotation on the Bloch sphere. In this paper, we restrict ourselves to rotations about axes that lie on the equatorial plane of the Bloch sphere, i.e. the $x{-}y$ plane, given by
\begin{eqnarray}\label{eq:rotation}
\hat U_{j}(\alpha)=\hat R_{\varphi}\left(\nu\alpha\right)&=&\ee{-i \nu\alpha(\cos(\varphi)\hat\sigma_{x}+\sin(\varphi)\hat\sigma_{y})/2}\\
&=&\cos\left(\frac{\nu\alpha}{2}\right)\hat\sigma_0-i\sin\left(\frac{\nu\alpha}{2}\right)\left(\cos(\varphi)\hat\sigma_{x}+\sin(\varphi)\hat\sigma_{y}\right)\,,
\end{eqnarray} 
where the index $j$ refers to a particular $\varphi$  and $\nu$, where $\varphi$ is the known angle between the axis of rotation and the $x{-}$axis and $\nu$ is real number that, along with $\alpha$, determines the rotation angle. As we see later, the $x{-}y$ plane restriction arrises naturally in the systems we consider. The measurement operators are defined as
\begin{equation}
\hat\mu_{j}(\alpha)=\hat U_{j}\dg(\alpha)\hat\mu_0\hat U_{j}(\alpha)\,.
\end{equation}
The relationship between the measurement statistics and the parameters which characterize the density matrix will be given by
\begin{eqnarray}\label{eq:abc}
n_{j}=\frac{\mathcal{N}_{j}}{2}\sum_{i=0}^{3} \mathrm{Tr}[\hat\mu_{j}(\alpha)\hat\sigma_{i}] \lambda_{i}\,.
\end{eqnarray}
By making an additional measurement, it is possible to solve the series of equations given by (\ref{eq:abc}) for $\lambda_{i}$ and $|\alpha|$. As before, measurements are chosen such that $\hat\mu_{j}(\alpha)$ generate a  linearly independent set of equations. If this condition is met, the equations yield two sets of solutions, corresponding to $\pm\alpha$, i.e. either the actual state or the phase-flipped version of the state. 

It is straightforward, but non-trivial, to extend this formalism to multiple-qubit systems. However this is not required as in practice one may use a maximum likelihood estimation method adapted for SCT to reconstruct the density matrix, which we will introduce in the next section.

\subsection{Maximum Likelihood}\label{sec:maxlike}
The solutions for $\lambda_{i}$ in (\ref{eq:2}) and (\ref{eq:abc}) are typically quite sensitive to fluctuations in measurement statistics, which can lead to unphysical density matrices. It is therefore favourable to use a maximum likelihood algorithm to reconstruct the density matrix.

Assuming the noise on the measurements has a Gaussian probability distribution, the probability of obtaining a set of $N$ measurement results $\{n_{j}\}$ is \cite{James2001}
\begin{eqnarray}\label{eq:begin}
P=A\prod_{{j}=0}^{N}\ee{-\frac{(n_{j}-\bar n_{j})^2}{2\sigma_{j}^2}}\,,
\end{eqnarray}
where $A$ is a normalization constant and $\sigma_{j}$ is the standard deviation of the ${j}$th measurement (given approximately by $\sqrt{n_{j}}$). The expectation value of each measurmeent is given by 
\begin{equation}
\bar n_{j}=\mathcal{N}_{j}\mathrm{Tr}[\hat\rho\hat\mu_{j}(\alpha)]\,,
\end{equation}
where $\hat\rho$ is the state we are trying to solve for.  To ensure a normalized, positive and hermitian density matrix, we restrict $\hat\rho$ to the form
\begin{eqnarray}
\hat\rho=\frac{\hat T^{\dagger}\hat T}{\mathrm{Tr}[\hat T^{\dagger}\hat T]}\,,
\end{eqnarray}
where
\begin{eqnarray}
T=\left(\begin{array}{cc}t_1 & 0 \\t_3+i t_4 & t_2\end{array}\right)\,,
\end{eqnarray}
for a single qubit and
\begin{eqnarray}
T_{2}=\left(\begin{array}{cccc}{t}_{1} & 0 & 0 & 0 \\{t}_{5}+i{t}_{6} & {t}_{2} & 0 & 0 \\{t}_{7}+i{t}_{8} & {t}_{9}+i{t}_{10} & {t}_{3} & 0 \\{t}_{11}+i{t}_{12} & {t}_{13}+i{t}_{14} & {t}_{15}+i{t}_{16} & {t}_{4}\end{array}\right)\,,
\end{eqnarray}
for a two qubit system. Maximising the likelihood that the physical density matrix $\hat\rho$ gave rise to the data $\{n_{j}\}$---i.e. maximising (\ref{eq:begin})---can be reduced to finding the minimum of the function 
\begin{eqnarray}
\mathcal L(\{{t}_{i}\},\alpha)=\sum_{{j}=0}^{N}\frac{(n_{j}-\bar n_{j})^2}{2\sigma_{j}^2}=\sum_{{j}=0}^{N}\frac{(n_{j}-\mathcal{N}_{j}\mathrm{Tr}[\hat\rho\hat\mu_{j}(\alpha)])^2}{2n_{j}}\,.
\end{eqnarray}

For large uncertainty in the measurement statistics, the likelihood function can have multiple local minima and the maximum likelihood algorithm may converge to a state and retardance that do not correspond to the actual state. This can be overcome given a priori knowledge of the purity of the state, or by taking more measurements. 

\section{Linear-optical implementation}\label{sec:exp}

For a linear-optical test of this method, we performed SCT on polarization-encoded one- and two-qubit states.

\subsection{Single qubits}
\begin{figure}[t]
\begin{center}
\includegraphics[width=\columnwidth]{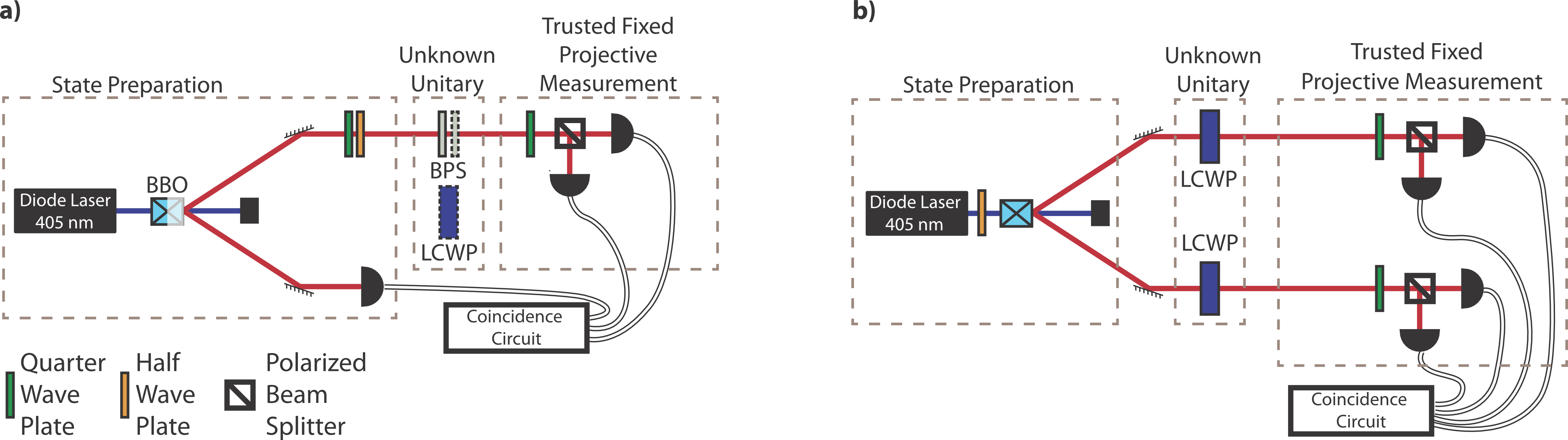}
\caption{Experimental scheme: a) A single-qubit state is prepared, unknown unitary operations are implemented using BPSs or an LCWP and the state is measured  using a trusted, fixed measurement; b) An entangled two-qubit state is prepared, unknown local unitary operations are implemented using LCWPs and the two-qubit state is measured using a trusted, fixed measurement. }
\label{fig:1}
\end{center}
\end{figure}

To prepare single-qubit states, we first pumped a $1mm$ long type-I down-conversion beta-barium borate (BBO) crystal with a $405nm$  diode laser, as shown in Fig. \ref{fig:1} a), which leads to spontaneous generation of pairs of photons. The detection of a horizontally polarized photon in one spatial mode of the down-converted photon-pair heralds the presence of a horizontally polarized photon in the other mode, i.e. the single qubit.  Using the method described in Bocquillon \emph{et al.} \cite{Bocquillon2009}, and based on an estimated average rate of 2-photon events in the signal conditional on a detected photon in the heralding arm of $0.96s^{-1}$, we calculated the $g^{(2)}(0)$ of our heralded single-photon source to be $g^{(2)}(0)=0.067$, indicating a high quality single-photon source. Quarter- and half-wave plates can then prepare the qubit in arbitrary polarization states.  

For the trusted, fixed projector, we chose $\hat\mu_0=\ket{R}{}\bra{R}{}$, implemented with a quarter-wave plate and a polarizing beamsplitter, followed by coincidence-counting, as shown in Fig. \ref{fig:1} a). To change the measurement basis, a unitary operation was applied to the state prior to measurement, implemented using either one or two identical pieces of a birefringent polymer sheet (BPS). The BPS is an optical element that alters the polarization state of light in the same way as a wave-plate, with the crucial difference that the retardance of the BPS is not known. The effect of the BPS on the state of the photon can be described by $\hat R_{\phi}(\alpha)$ given in (\ref{eq:rotation}), where $\alpha$ is the retardance of the BPS, $\varphi$ is the alignment of the BPS's optical axis with respect to the polarization of the light, and $\hat\sigma_{i}$ are in the $\ket{R}{}/\ket{L}{}$ basis. Using different alignments of the BPS, we constructed the following operators $\hat\mu_{j}(\alpha)=\hat U_{j}(\alpha)\dg\hat\mu_0\hat U_{j}(\alpha)$ using different implementations of $\hat U_{j}(\alpha)$: 
\begin{eqnarray}
\hat U_0(\alpha)&=&\hat \sigma_{0}\,,\\
\hat U_1(\alpha)&=&\hat R_{0}(\alpha)\,,\\
\hat U_2(\alpha)&=&\hat R_{\frac{\pi}{2}}(\alpha)\,,\\
\hat U_3(\alpha)&=&\hat R_{\pi}(2\alpha)\,,\\
\hat U_4(\alpha)&=&\hat R_{\frac{3\pi}{2}}(2\alpha)\,.
\end{eqnarray}

Note that rotations of $2\alpha$ were required to satisfy the linear-independence condition on (\ref{eq:2}). These rotations were realized with a double-pass through the BPS---the sheet was cut in half and the pieces placed in succession. The beam passed through sections of the pieces that were within close proximity before partition, to ensure consistency in the birefringence. One piece was then removed to perform rotations of $\alpha$, before removal of the remaining piece to measure $\hat\mu_0$.  This measurement sequence  eliminated the need for realignment between measurements.

In this particular linear-optical implementation of the scheme, both output ports of the PBS are easily accessible. We therefore measured the complimentary projectors, $\hat\mu_{j}'(\alpha)=\hat U_{j}(\alpha)\dg\hat\mu_0'\hat U_{j}(\alpha)$ where $\hat\mu_0'=\ket{L}{}\bra{L}{}$, to determine $\mathcal{N}_{j}$.

Using this method, SCT was performed on 14 states: six that lie at the ends of the three axes and eight other states defined by all combinations of $\theta=\pm\frac{\pi}{4}$ and $\phi=\{\pm\frac{\pi}{4},\pm \frac{3\pi}{4}\}$ where $\ket{\psi}{}\nobreakspace=\nobreakspace\cos\left(\frac{\theta}{2}\right)\ket{R}{}\nobreakspace+\nobreakspace\ee{i \phi}\sin\left(\frac{\theta}{2}\right)\ket{L}{}$. Each state was reconstructed using the maximum-likelihood estimation method described in Section \ref{sec:maxlike} in conjunction with the YALMIP solver \cite{Lofberg2004}. 

\begin{figure}[t]
\begin{center}
\includegraphics[width=\columnwidth]{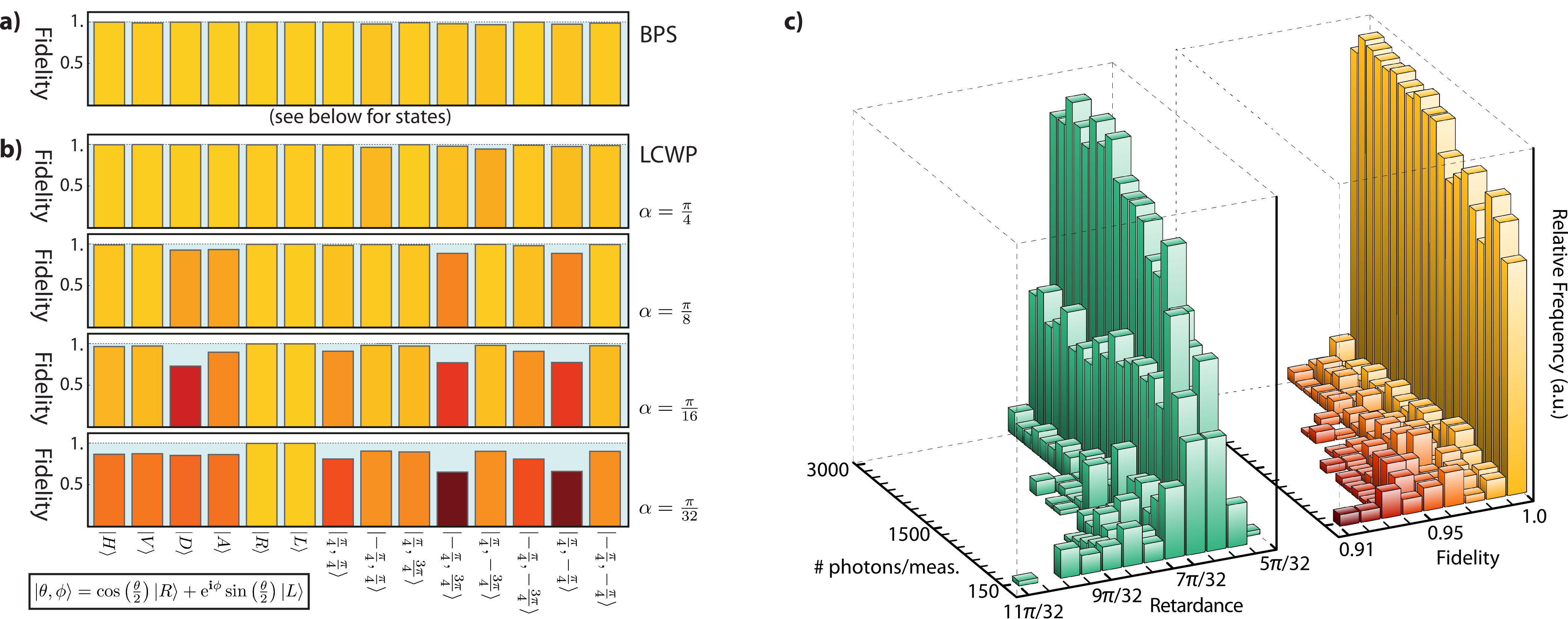}
\caption{Single-shot fidelity between SCT and ST for 14 states distributed on the Bloch sphere, using: a) a birefringent polymer sheet (BPS) as the uncalibrated wave plate; and b) using a liquid crystal wave plate (LCWP) set to different values of the retardance $\alpha$. c) Histogram showing the predicted retardance and the fidelity between SCT and ST, for 100 runs of SCT with varying numbers of total photons collected (and hence levels of counting error). All used $\ket{\psi}{}=\ket{H}{}$ and the BPS.}
\label{fig:2}
\end{center}
\end{figure}

For comparison, standard tomography (ST) was also performed on each of the states.  ST required fewer projectors $\hat\mu_{j}=\hat U_{j}\dg\hat\mu_0\hat U_{j}$, given by the unitary operators:
\begin{eqnarray}
\hat U_0&=&\hat\sigma_{0}\\
\hat U_1&=&R_{0}\left(\frac{\pi}{2}\right)\\
\hat U_2&=&R_{\frac{\pi}{2}}\left(\frac{\pi}{2}\right)\,,
\end{eqnarray}
and their complimentary projectors, $\hat\mu_{j}'=\hat U_{j}\dg\hat\mu_0'\hat U_{j}$ where $\hat\mu_0'=\ket{L}{}\bra{L}{}$. We used the fidelity $F=\mathrm{Tr}[\sqrt{\sqrt{\hat\rho_{\textsc{st}}}\hat\rho_{\textsc{sct}}\sqrt{\hat\rho_{\textsc{st}}}}]^2$ to characterize the success of SCT, where $\hat\rho_{\textsc{st}}$ and $\hat\rho_{\textsc{sct}}$ are the density operators reconstructed using ST and SCT respectively. We achieved an average fidelity $\bar F=0.99\pm0.01$ and determined the retardance of the BPS (at $405nm$) to be $\alpha=0.58 \pm 0.03$ (close to $\pi/6$). Individual fidelities for each state are shown in Fig. \ref{fig:2} a).

\subsection{Dependence on retardance}

To determine the effectiveness of SCT as a function of the retardance, we repeated the experiment using BolderVision Optik liquid crystal wave plates (LCWP), as shown in Fig. \ref{fig:1} a). In an LCWP, the birefringence is a function of the applied voltage, allowing for the retardance to be set arbitrarily. A single LCWP was used to implement rotations of $\alpha$ and $2\alpha$ using different voltage settings. Fidelities for different setting of $\alpha$ are shown in Fig. \ref{fig:2} b). The average fidelity decreases with the size of the retardance. The fidelity for states $\ket{\psi}{}=\ket{R}{}$ and $\ket{\psi}{}=\ket{L}{}$ always remains close to unity. This is a consequence of our choice of measurement projectors. Using the LCWP, we find a slightly lower average fidelity with a greater spread over the different states, even for values of $\alpha$ larger than that of the BPS. We attribute this to errors associated with the non-linear dependence of the retardance on the voltage passed through the LCWP. This influences how accurately $2\alpha$ can be set with respect to $\alpha$.

\subsection{Dependence on noise}

\begin{figure}[t]
\begin{center}
\includegraphics[width=\columnwidth]{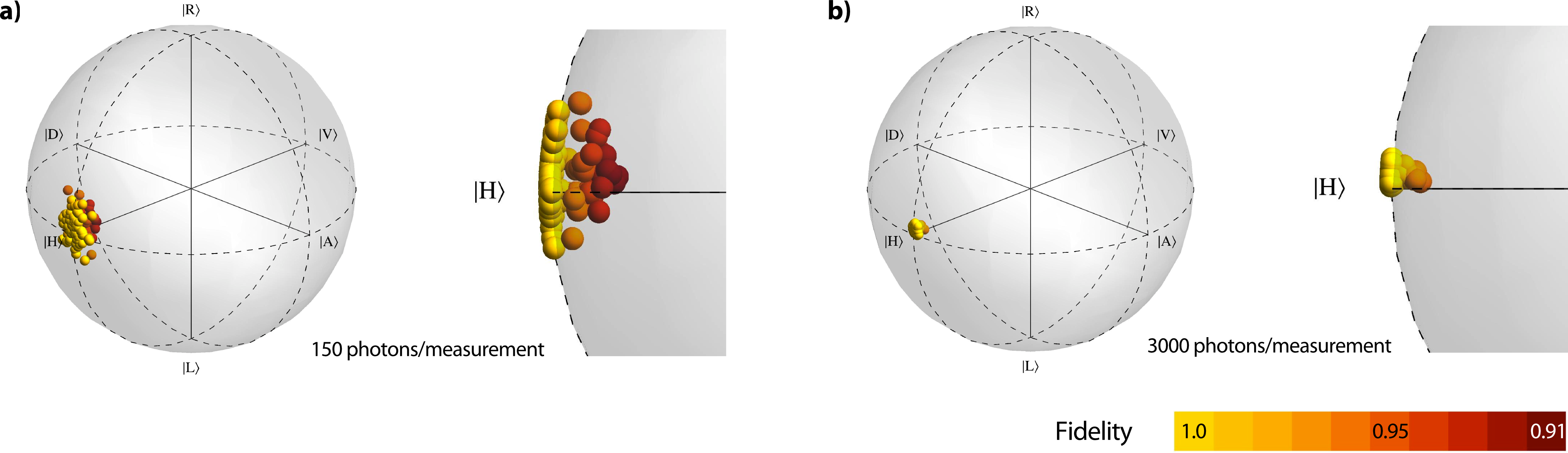}
\caption{Distribution of states on the Bloch sphere as predicted by SCT for: a) high noise; and b) low noise. Notice the darker red cluster of states in Fig. a), which corresponds to a local minimum in the likelihood function. There are 100 states shown in each sphere.  }
\label{fig:4}
\end{center}
\end{figure}

 To  investigate the dependence of the fidelity on the amount of noise, we performed SCT on the state $\ket{\psi}{}=\ket{H}{}$ using a retardance $\alpha=\pi/32$. The noise scales inversely with the square root of the number of counts and by collecting from 150 to 3000 photons we varied the noise between roughly $1\%$ and $10\%$. For each noise setting, SCT was repeated 100 times. The resulting fidelities and predicted retardances are presented in Fig. \ref{fig:2} c). Notice that the majority of predicted fidelities and retardances correspond to the actual values. However, recall that  the likelihood function can have multiple local minima, making it possible for the maximum likelihood algorithm to converge to a state and retardance that do not correspond to the actual state. This shows up in the high-noise, or low photon-number, region and can be seen as a small bump in the fidelity histogram (and to a lesser extent in the retardance histogram) in Fig. \ref{fig:2} c).  As the noise decreases, only high-fidelity results corresponding to the actual state and retardance remain. Fig. \ref{fig:4} shows the distribution of states on the Bloch sphere as predicted by SCT for high and low noise  amounts. The darker red cluster of states in Fig. \ref{fig:4} a) also correspond to a local minimum in the likelihood function. 
 
\subsection{Entangled qubits}

 For multi-qubit tomography, entangled two-qubit states, $\ket{\psi}{}=a\ket{HH}{}+b\ket{VV}{}$, were prepared using two $1mm$ long type-I down-conversion BBO crystals with optical axes aligned in perpendicular planes \cite{Kwiat1999a} pumped by a $405nm$ diode laser, as shown in Fig. \ref{fig:1} b). The parameters $a$ and $b$ which characterise the state were tuned by changing the polarization of the pump beam with a HWP. LCWPs in each arm, implemented the unknown unitary operations, followed by detection of $\hat\mu_0=\ket{R}{}\bra{R}{}$ in both modes. For this two-qubit experiment, LCWPs were chosen over the BPSs due to realignment issues that accompany the insertion (as opposed to removal) of optical elements in the beam-path. The concurrence \cite{Wootters1998} for states reconstructed using SCT and ST, as well as the fidelity between the two, are shown for a variety of entangled states in Table \ref{tab:1}. 
\begin{table}[h]
\begin{center}
\begin{tabular}{ccc}
\hline
\hline
$C_{\textsc{SCT}}$&$C_{\textsc{ST}}$&$F$~~~~~\\
\hline
\hline
$0.905\pm0.005$&$0.927\pm0.003$&$0.968 \pm 0.0028$~*~~\\
$0.566\pm0.005$&$0.562\pm0.005$&$ 0.978 \pm 0.0012$~~~~~\\
$0.337\pm0.005$&$0.328\pm0.004$&$0.977 \pm 0.0032$~~~~~\\
$0.004\pm0.004$&$0.010\pm0.003$&$0.989 \pm 0.0015$~**\\
\hline
\end{tabular}
\end{center}
\caption{The concurrence \cite{Wootters1998} as predicted by both tomographic techniques and fidelity between the two reconstructions for states with different degrees of entanglement. Density matrices for * and ** are shown in Figure \ref{fig:5} a) and b) respectively. Error bars were determined by using Poissonian noise in a Monte Carlo simulation}
\label{tab:1}
\end{table}

\begin{figure}[t]
\begin{center}
\includegraphics[width=\columnwidth]{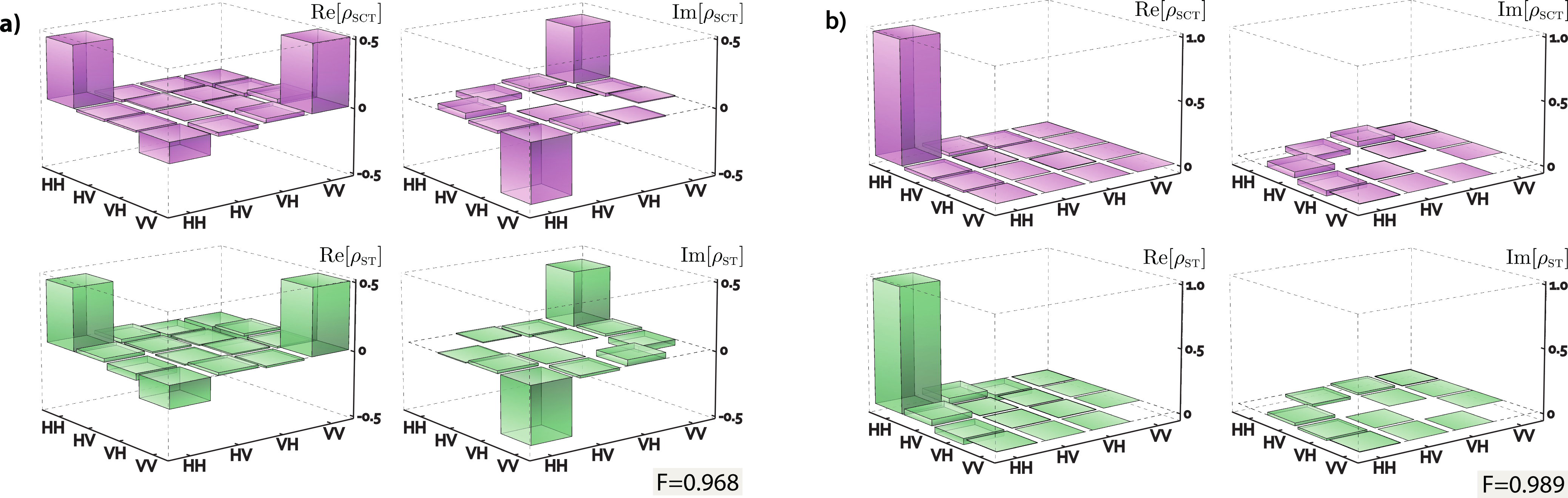}
\caption{Comparison between density matrices reconstructed using SCT and ST: a) $F=0.968 \pm 0.0028$;  b) $F=0.989 \pm 0.0015$. For concurrences and other fidelities, refer to Table \ref{tab:1}.}
\label{fig:5}
\end{center}
\end{figure}

 Fig. \ref{fig:5} shows density matrices for the most- and least-entangled states. The retardances were determined to be $\alpha_1=0.810 \pm 0.009$ and $\alpha_2=0.760 \pm 0.006$, compared with the known value $\alpha=\pi/4\approx 0.785$.

\section{Potential applications to photosynthetic systems}\label{sec:photo}

In this section, we give some detail about how this technique may be applied to performing quantum state tomography on biological molecules, such as chromophores in photosynthetic systems, where the transition dipole moment and its orientation are not known with high accuracy. 

\subsection{Unitary operation}

Consider a molecule with two states of interest---the ground state $\ket{g}{}$ and the excited state $\ket{e}{}$---that undergoes an interaction with an EM field, e.g. a laser pulse, given by the interaction Hamiltonian
\begin{eqnarray}
\hat H=\hat{\mathbf{d}}\cdot E(t)= \vec{d}_{eg}\cdot  E(t) \ket{e}{}\bra{g}{}+\vec{d}_{ge}\cdot  E(t)^* \ket{g}{}\bra{e}{}\,,
\end{eqnarray}
where $E(t)$ is the electric field, $\hat{\mathbf{d}}$ is the transition dipole moment operator and $\vec{d}_{eg}=\bra{e}{}\hat{\mathbf{d}}\ket{g}{}=d^*_{ge}$. An ultra-fast pulse with a carrier frequency $\omega_{c}$ can be described by $E(t)=\mathcal{E}(t)\ee{i\omega_{c}{t}}\vec{e}+\mathrm{h.c.}$, 
where $\mathcal{E}(t)$ is the temporal profile of the pulse and $\vec{e}$ is the polarization vector. 
Under the rotating wave approximation, we can write the Hamiltonian in the interaction picture as
 \begin{eqnarray}
\hat H=(\vec{d}_{eg}\cdot  \vec{e})\mathcal{E}(t)\ee{i(\omega_{c}-\omega_{e})t} \ket{e}{}\bra{g}{}+(\vec{d}_{eg}\cdot  \vec{e})^*\mathcal{E}(t)^*\ee{-i(\omega_{c}-\omega_{e})t}\ket{g}{}\bra{e}{}\,,
\end{eqnarray}
where $\omega_{e}$ is the $\ket{g}{}\rightarrow\ket{e}{}$ transition energy. The unitary governing the interaction is given by
\begin{eqnarray}\label{eq:unitary4}
\hat U_{j}(\alpha)=\ee{-\frac{i}{\hbar}\int dt \hat H(t)}=\ee{-i\left(\mathrm{Re}[\Omega]\hat\sigma_{x}+\mathrm{Im}[\Omega]\hat\sigma_{y}\right)/2}=\ee{-i\nu\alpha\left(\cos(\varphi)\hat\sigma_{x}+\sin(\varphi)\hat\sigma_{y}\right)/2}\,,
\end{eqnarray}
where $\Omega=\frac{2}{\hbar}(\vec{d}_{eg}\cdot  \vec{e})\int dt\mathcal{E}(t)\ee{i(\omega_{c}-\omega_{e})t}$ and
\begin{eqnarray}\label{eq:angle}
\nu&=&\frac{2}{\hbar}\left|\int dt\mathcal{E}(t)\ee{i(\omega_{c}-\omega_{e})t}\right|\,,\\
\alpha&=&(\vec{d}_{eg}\cdot  \vec{e})\,,\\
\varphi&=&\mathrm{arg}\left(\int dt\mathcal{E}(t)\ee{i(\omega_{c}-\omega_{e})t}\right)\,.
\end{eqnarray}

Comparing with (\ref{eq:rotation}), we can see that (\ref{eq:unitary4}) constitutes a rotation of angle $\nu\alpha$, where $\varphi$---determined by the relative phase of $\mathcal{E}(t)$---defines the rotation axis. Notice from (\ref{eq:angle}) that $\nu$ is linearly proportional to the square root of the intensity of the pulse at the transition frequency. We can therefore execute rotations of $2\alpha$ by changing the intensity of the pulse, without any knowledge of $\vec{d}_{eg}\cdot  \vec{e}$.

\subsection{Projective measurement}

We can perform a measurement corresponding to the operator
\begin{eqnarray}
\hat\mu_{0}=\ket{e}{}\bra{e}{}\,,
\end{eqnarray}
by preparing multiple copies of the state and performing statistics on the number of photons emitted. A molecule in the excited state will---on a long enough time-scale---emit a photon, while one in the ground state will not; a molecule in a superposition of ground and excited states $a\ket{g}{}+b\ket{e}{}$ will emit a photon with probability $|b|$. The measurement statistics will be given by
\begin{equation}
n_{j}=\mathcal{N}_{j}\mathrm{Tr}[\hat U_{j}(\alpha)\hat\rho\hat U_{j}(\alpha)\dg\hat\mu_0]\,,
\end{equation}
where $\hat U_{j}(\alpha)$ is given in (\ref{eq:unitary4}) and $\mathcal{N}_{j}$ depends on the solid angle of the detector, the position of the detector with respect to the dipole moment of the molecule and losses. 

Measurement statistics can be collected either by repeated preparation and measurement of a single molecule, recently demonstrated by Hildner \emph{et al.} \cite{Hildner2011a,Hildner2011}, which is analogous to the linear-optical implementation discussed above; or by preparing an ensemble of molecules in the same state and measuring the fluorescence of the sample, where the time-integrated intensity of the emitted light is directly proportional to the number of photons.  An open question is whether it will be possible for the self-calibrating technique to compensate for an ensemble of randomly oriented molecules. This is currently  being investigated.  

In principle, the extension to multiple entangled molecules should be straightforward: even if individual molecules can not be spatially isolated, independent rotations on each qubit could be executed with appropriately shaped pulses, as long as there are different transition energies for each molecule.  However, in the case of coupled molecules, there are technical difficulties associated with addressing the transition energies of the individual molecules---the system rather absorbs light at the eigen-energies of the coupled Hamiltonian. In this case, a system of two coupled two-level systems can be approximated as a three-level system (where the fourth level is typically disregarded). Self-calibrating quantum state tomography of coupled two-level system will require an extension of our formalism to a three-dimensional Hilbert space. This is also currently being investigated.

\section{Conclusion}

In conclusion, we have introduced and demonstrated a technique for performing quantum state tomography of multi-qubit systems which does not rely on complete knowledge of the unitary operation used to change the measurement basis. Our technique characterizes the state of interest with high fidelity as well as recovering the unknown unitary operation. We find that in the context of QST of polarization-encoded qubits, it is possible to do away with a well-calibrated half-wave plate and replace it with an inexpensive uncharacterized piece of birefringent material. 

A future extension of this work will attempt to incorporate unknown rotation axes and non-unitary transformations, such as those that introduce noise to the system, thereby reducing the purity of the state by an unknown amount. 

We anticipate our method to be applicable to quantum state tomography of systems that do not have well-characterized transition dipole moments, such as biological molecules like those found in photosynthetic systems, as well as systems that experience variability during the fabrication process, such as colloidal quantum dots.

AMB, AD and DFVJ thank DARPA (QuBE) for support. DHM, LAR and AMS thank NSERC, CIFAR and QuantumWorks for support. We also thank Xingxing Xing for the serendipitous discovery of the birefringence of his smartphone screen protector. 



\end{document}